\documentclass[preprintnumbers,amsmath,amssymb,aps]{revtex4}

\usepackage{graphicx}
\usepackage{multirow}
\usepackage{tabularx}

\preprint{HU-EP-07/63}
\preprint{SFB/CPP-07-85}
\preprint{COLO-HEP-532}

\newcommand{\bee}{\begin{equation}}
\newcommand{\ee}{\end{equation}}
\newcommand{\beea}{\begin{eqnarray}}
\newcommand{\eea}{\end{eqnarray}}

\begin{document}

\title{Topological susceptibility in two-flavor QCD}
\author{Thomas DeGrand}
\affiliation{
Department of Physics, University of Colorado,
Boulder, CO 80309 USA
}
\author{Stefan Schaefer}
\affiliation{
Institut f{\"u}r Physik, Humboldt Universit\"at zu Berlin, Newtonstr.~15,  12489 Berlin, Germany
}

\begin{abstract} 
We compute the topological susceptibility  in QCD with two flavors of dynamical fermions using
numerical simulation with overlap fermions.
\end{abstract}
\maketitle

\section{Introduction}
Arguably, the quantity in QCD which is most sensitive to the number of flavors and the masses of dynamical
fermions is the topological susceptibility $\chi_T$, which is expected
\cite{Crewther:1977ce,DiVecchia:1980ve,Leutwyler:1992yt}
 to vanish at small quark mass $m_q$ as
\bee
\chi_T = \frac{m_q \Sigma}{N_f}.
\label{eq:lschpt}
\ee
($\Sigma$ is the condensate; $N_f$ the number of flavors.)
As the masses of the dynamical fermions rise, the naive expectation is that
 $\chi_T$ also rises and saturates
at its quenched value $\chi_Q$. This behavior is encoded in the large-$N_c$ formula of di Vecchia and Veneziano and
of Leutwyler and Smilga\cite{DiVecchia:1980ve,Leutwyler:1992yt},
\bee
\frac{1}{\chi_T} = \frac{N_f}{m_q\Sigma} + \frac{1}{\chi_Q}.
\label{eq:durr}
\ee

Also arguably, the topological susceptibility is also the quantity which in lattice 
simulations is most sensitive
to lattice artifacts. The situation before, say, 2001, based on simulations
using fermions which did not respect exact chiral symmetry at nonzero lattice spacing,
was murky (compare the figures in  Ref. \cite{Durr:2001ty}).
Even today  \cite{Bernard:2003gq}, simulations with improved lattice actions, which still do not encode
information about the index theorem and the anomaly, do not give a crisp realization
of Eq. \ref{eq:lschpt}.
In the last few years, with the advent of lattice discretizations  of the Dirac operator
 (specifically overlap fermions\cite{Neuberger:1997fp,Neuberger:1998my})
 which preserve
 full continuum chiral symmetry \cite{Ginsparg:1981bj}, this situation has changed.
We expect that such dynamical fermions will possess enough symmetry to realize
 Eq. \ref{eq:lschpt} automatically, by suppressing the production of topology at small
quark mass.
With these lattice actions, it is also very simple to assign a topological charge to 
a particular gauge configuration through the index theorem: one can just count the number of 
zero modes of the Dirac operator. Recently a number of 
studies \cite{Fodor:2004wx,DeGrand:2005vb,Egri:2005cx,Aoki:2007pw} of the topological
susceptibility with two flavors of dynamical fermions have appeared. This paper
continues the story.

One defect that all lattice simulations possess is that it is very difficult to 
move from one topological
sector to another during the Markov evolution which generates the data set.
This happens because all simulations replace the fermionic contribution to the action by a noisy
estimator\cite{Gottlieb:1987mq}. At a topological boundary in the space of gauge 
configurations, this noisy estimator
tends to overestimate the barrier height against tunneling. 

One strategy to overcome this is to do simulations in sectors of fixed topology.
Two variations on this idea are that of
Egri, {\it et al.} \cite{Egri:2005cx}, who extract $\chi_T$ from the ratio of the
partition function in topological sectors along a boundary, and of
 S.~Aoki, {\it et al.} \cite{Aoki:2007pw}, who compute $\chi_T$ from the
asymptotic behavior of a pseudoscalar
correlation function measured in gauge field backgrounds of fixed topology.

We adopt a somewhat simpler approach: we do simulations with an algorithm which is tuned
to maximize the tunneling rate among topological sectors. 
While at the end of the day
we are not happy with our tunneling rates, this direct approach does seem to have 
been successful. We observe a linear dependence of the topological susceptibility on the quark mass,
which is consistent with Eq. \ref{eq:lschpt} when it is combined with our previous,
more direct measurements of the condensate \cite{DeGrand:2007tm}.

We outline the rest of the paper: in the next section we describe the simulations.
We then pause to describe a method \cite{Blum:2001xb,Aoki:2005ga,Allton:2007hx,Aubin:2007pt}
for performing spectroscopy calculations in small volumes
(needed to plot $\chi_T$ vs pseudoscalar mass). Sec. IV then contains a description of 
our data and analysis of the topological susceptibility. We give our conclusions in Sec. V.

\section{Simulations}

We performed simulations in two-flavor QCD using overlap fermions.
Our data set uses a lattice volume of $12^4$ points. 
The overlap operator uses a ``kernel action'' (the nonchiral
 action inserted in the usual overlap formula) with nearest and
 next-nearest (diagonal) neighbors. The gauge connection is the
 differentiable hypercubic smeared link of
 Ref. \cite{Hasenfratz:2007rf}. 
Details of the actions are described in
 Refs.
\cite{DeGrand:2000tf,DeGrand:2004nq,DeGrand:2006ws,DeGrand:2006nv,DeGrand:2007tm}.

We employ the reflection/refraction algorithm
first devised in Ref.~\cite{Fodor:2003bh}. In order to improve the tunneling
rate and precondition the fermion determinant we use one or two additional heavy 
pseudo-fermion fields as suggested by Hasenbusch\cite{Hasenbusch:2001ne}.
The integration is done with multiple-time scales\cite{Urbach:2005ji}. The
 runs 
for all sea quark masses were performed within a few months 
on a cluster of 32 Opteron CPU's which are connected by an Infiniband network.
We compute eigenvalues using the ``Primme'' package of
McCombs and Stathopoulos\cite{primme}.

\section{Supporting calculations}

\subsection{Lattice spacing}
We determine the lattice spacing from a fairly standard measure of the Sommer parameter
\cite{Sommer:1993ce}. Its value at our three dynamical quark masses
 is summarized in Table \ref{tab:R0}.

\begin{table}
\begin{tabular}{c|cc}
\hline
$am_q$      &     $r_0/a$     & $am_{ps}$   \\
\hline
0.03  \ \  & \ 3.70(5)  \   & 0.324(10)  \\
0.05  \ \  & \ 3.49(4)  \   & 0.430(6)   \\
0.10  \ \  & \ 3.39(3) \    & 0.589(6)   \\
\end{tabular}
\caption{\label{tab:R0} Sommer parameter $r_0$ and
 pseudoscalar mass as a function of dynamical fermion mass. }
\end{table}

\subsection{Spectroscopy on small lattices}
To show the mass dependence of $\chi_T$, it is useful to measure the mass
of the pseudoscalar meson. Spectroscopy done with valence quarks with the same
(antiperiodic)
temporal boundary conditions as for the sea quarks  gives rise to
meson correlators which are periodic in the temporal variable $t$, and the
maximum separation of source and sink meson correlators is $t=T/2$ if the
lattice has temporal extent $T$. This is uncomfortably small if $T=12$.

We can effectively double $T$ by using a trick we learned from N. Christ, which
has been used by several groups for computing weak matrix 
elements \cite{Blum:2001xb,Aoki:2005ga,Allton:2007hx,Aubin:2007pt}:
Take a valence Dirac operator with periodic temporal boundary conditions and
compute its propagator, $S_P(x)$ (we assume a source at $t=0$ for simplicity).
Take a second valence Dirac operator with antiperiodic temporal boundary
conditions, and compute its propagator $S_A(x)$. Now add the propagators to produce
 $S_{P+A}=(S_P(x)+S_A(x))/2$, and use this propagator
to construct hadron correlators. The resulting correlator will be a hyperbolic
cosine with midpoint at $t=T$ (see Eq. \ref{eq:PPA}, below). This is called the ``P+A trick.''
In the context of chiral perturbation theory, and in the p-regime, this is a completely
legitimate way to compute low energy coefficients and processes involving one hadron in the
initial and/or final state. 
The demonstration that this is so is a simple variation on work of Sachrajda and Villadoro\cite{Sachrajda:2004mi}.

We follow them and imagine that we have some ``fiducial'' boundary conditions, imposed on the sea quarks,
 (Sachrajda and Villadoro,
who are thinking about observables with nonzero spatial momentum, take these to be periodic spatial ones)
and some ``twisted'' boundary conditions, which are obeyed by some of the valence quarks.
 For us the fiducial boundary conditions are antiperiodic in 
time and the twisted ones are periodic in time. The twisted quark fields
are redefined through
\begin{equation}
q(x)\equiv V(x)\,\tilde q(x) \qquad\textrm{where}\qquad
V(x)\equiv\exp (i\frac{\Theta_\mu}{L_\mu}x_\mu).
\end{equation}
(a single sum on $\mu$ is implied)
where $\Theta_\mu$ is the rotation needed to turn $q(x)$ into  $\tilde q(x)$ which obeys the fiducial boundary conditions
(periodic in space for Sachrajda and Villadoro,
antiperiodic in time for us). $L_\mu$ is the length of the simulation volume in direction $\hat \mu$.
Expressed in terms of the fiducial fields, the twisted Dirac operator is defined as $\tilde D_\mu = D + iB_\mu$, where $B_\mu=\Theta_\mu/L_\mu$, 
such that the QCD Lagrange density reads ${\cal L}=\bar{\tilde q}(x) \tilde D \tilde q(x)$. 
This is QCD in the presence of an external vector field, coupling to quarks, with charges 
determined by the phases of the boundary conditions. 

Now for the chiral Lagrangian. The composite field in ${\cal L}_{eff}$, 
$U$, is a matrix  which satisfies the boundary condition
\begin{equation}
U(x_i+L)= V_i U(x_i) V_i^\dagger\,.
\end{equation}
So again,  the relation to its fiducial value is given by
\begin{equation}
 U(x)\equiv V(x) \tilde U(x)V^\dagger(x) \,.
\end{equation}
The vector field turns the ordinary derivative in the chiral Lagrangian into a covariant derivative,
\begin{equation}
\tilde D_\mu \tilde U = \partial_\mu \tilde U + i[B_\mu,\tilde U].
\end{equation}
The $B$ field introduces an extra boost to the momentum
of the meson, according to its flavor content: in a quark basis, where $i$ and $j$ label
the flavors of the quarks,
$B_{ij}= \theta_i - \theta_j$.
For us, this is the zeroth component of the four vector $B_\mu$, the other components are zero.
Thus a  meson made of two periodic quarks, or one made of two antiperiodic
quarks, is periodic ($B=0)$; if it is made of one of each kind, $B=\pi/T$.
Because  of the way twist enters the chiral Lagrangian, low energy constants are not affected by it\cite{Sachrajda:2004mi}.

The P+A trick is a very mild form of partial quenching (PQ). The valence and the sea quarks are given
the same mass; they differ only in the boundary condition.
In a usual PQ simulation, one looks at correlators of valence quarks separately.
Here, we want to combine the two boundary condition quarks, twisted and untwisted,
into one and compute its correlator.
That amounts to taking a sum of correlators of mesons made of ordinary and valence quarks:
\begin{equation}
C(t) = \frac{1}{4}(C_{vv}+C_{vs}+C_{sv}+C_{ss})=\frac{1}{2}(C_P +C_A);
\end{equation}
P and A mean periodic or antiperiodic. The former collects the $vv$ and $ss$ terms,
the latter the cross correlators. How does such a correlator look?
The free field space averaged correlators in a box of length $T$ are
\begin{equation}
C(t) = \sum_{l=-\infty}^{l=\infty} \frac{e^{i\omega_l t}}{\omega_l^2+m^2}
\end{equation}
where $\omega_l = 2\pi l/T$ for periodic and $\omega_l = (2l+1)\pi /T$
for antiperiodic temporal boundary conditions,
so
\begin{equation}
C_P(t)=\frac{\cosh(m(T/2 -t))}{2m\sinh(mT/2)},
\end{equation}
\begin{equation}
C_A(t)=\frac{\sinh(m(T/2 -t))}{2m\cosh(mT/2)},
\end{equation}
and
\begin{equation}
\frac{C_P(t)+C_A(t)}{2}= \frac{\cosh(m(T-t))}{2m\sinh(mT)}
\label{eq:PPA}
\end{equation}
peaks at $t=0$ while
\begin{equation}
\frac{C_P(t)-C_A(t)}{2}= \frac{\cosh(mt)}{2m\sinh(mT)}
\end{equation}
peaks at $t=T$.

Because we take degenerate masses, our version of PQ chiral perturbation theory does
not contain the usual PQ double-pole artifacts.
What will be different are the finite volume corrections to the chiral logarithms.
If high accuracy is needed, they can be taken from the paper of
Sachrajda and Villadoro, except $\exp(-mL)$ is replaced by   $\exp(-mT)$ and some combinatorial factors
must be corrected. In many simulations (but not ours), $T\gg L$ and these corrections
are negligible. If we were trying to achieve great accuracy, we would have to include them.
For present purposes, we neglect them.

\begin{figure}
\begin{center}
\includegraphics[width=0.8\textwidth,clip]{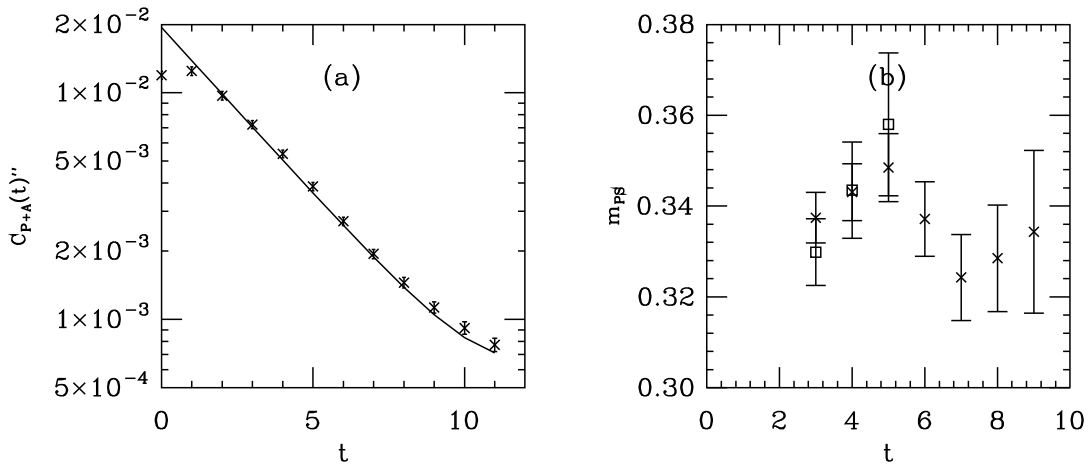}
\end{center}
\caption{(a) Range fit to ``$P+A$'' pseudoscalar correlator,
$am_q=0.03$. (b) Range fits, from $t$ to $t=10$ from the $P+A$ data set 
are shown in crosses.
Squares are the fits to ordinary wall source to point sink correlators,
with antiperiodic time boundary conditions.
\label{fig:pastuff}
}
\end{figure}

As an example of how well the P+A trick works, see Fig. \ref{fig:pastuff}.
We show a correlator and its fit, and the resulting masses from a set of range fits,
 from a 40 lattice subset of our $am_q=0.03$
data set. We use this methodology to compute
the pseudoscalar masses shown in Table \ref{tab:R0}. 

\section{Results}
We now turn to the topological charge itself. At each of the three quark masses, we 
attempted (roughly) to optimize the simulation parameters to enhance tunneling.
At $am_q=0.03$ and  $am_q=0.05$  we used three pairs of 
Hasenbusch pseudofermions. At the $am_q=0.10$ we took two pairs.
 Most of the $am_q=0.05$ runs and a small fraction of the other masses used
 trajectory length of one half time unit, and the bulk of the running at $am_q=0.03$ and 0.10
 running used trajectories of unit length.
We ran at about an 80 per cent acceptance rate for all three quark masses.

\begin{figure}
\begin{center}
\includegraphics[width=0.278\textwidth,clip,angle=-90]{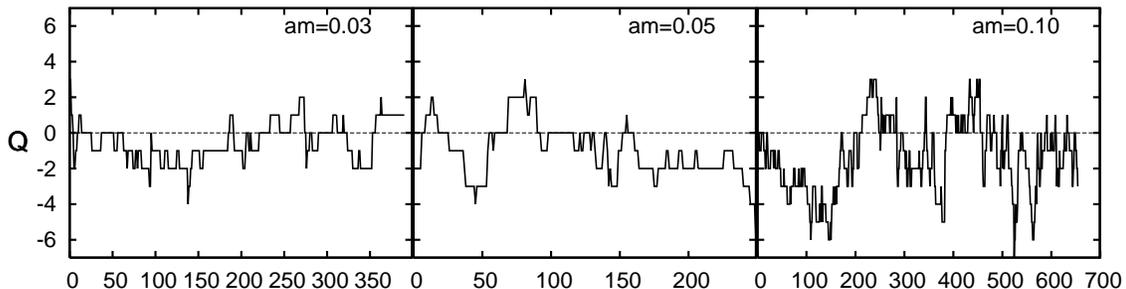}
\end{center}
\caption{
Time history of the topological charge in units of molecular dynamics trajectories.
\label{fig:qhistory}
}
\end{figure}

\begin{figure}
\begin{center}
\includegraphics[width=0.31\textwidth,angle=-90,clip]{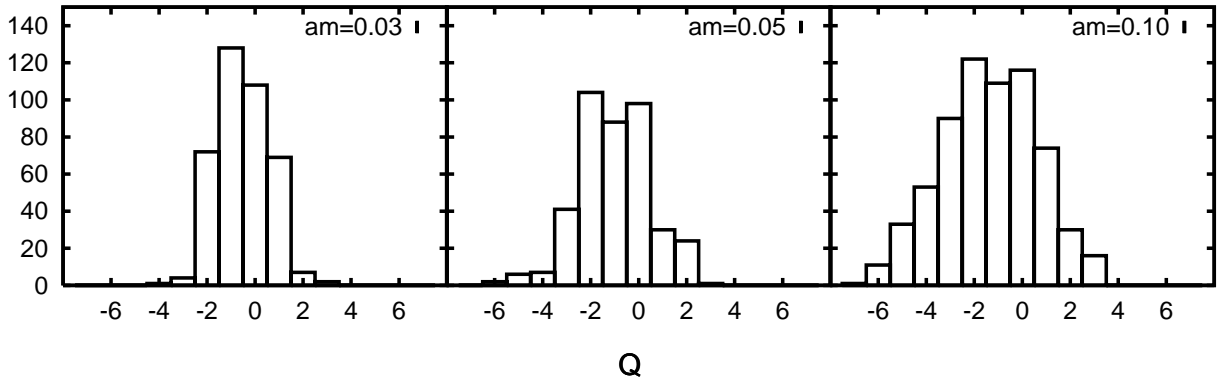}
\end{center}
\caption{
\label{fig:qhist}
Histogram of topological charge for the three sea quark masses. The
topology is measured after each trajectory, so auto-correlations are not
taken into account.
}
\end{figure}

Histories  of the topological charge are shown in Fig. \ref{fig:qhistory}.
Without any more analysis, one can see immediately that topological changes do occur relatively
frequently. That is good. However, one also sees the presence of long period
variations, probably longer than our simulation time.
This is revealed most clearly in histograms of the topological charge, Fig. \ref{fig:qhist}:
the distributions are not symmetric about the origin.

These long fluctuations are a severe short-coming of our analysis. In order to get reliable
errors, we tried to measure
the auto-correlation times $\tau_{\rm int}$ of $Q$ using the methods described in Ref.~\cite{Wolff:2003sm}.
Of course, this analysis is not going to be sensitive to time auto-correlations which are 
long compared to the total simulation time. 
The results are displayed in Table~\ref{fig:AC}. It shows that within $2\sigma$, the
 topological charge averages to zero. To take the effect of the
auto-correlations on the error estimate  into account, we multiplied the naive errors by
$\sqrt{2\tau_{\rm int}}$. 
The measurement of the integrated auto-correlation time, however, is quite unreliable.
Contrary to expectation, its averages decrease with the quark mass (although
this statement has no statistical significance). This is probably
only due to the fact that with the higher statistics of the $am_q=0.1$ ensemble,
there is actually the chance to be sensitive to longer correlations.

The relevant quantity for the susceptibility is $\langle Q^2\rangle$.
To  estimate the systematic error on the extraction of the susceptibility 
from the fact that $Q$ does not average to zero, we also computed
$\langle Q^2 -\langle Q \rangle ^2 \rangle$. The measured values are
also given in Table~\ref{fig:AC}. Both quantities should agree 
in the limit of infinite statistics. It turns out that they have errors of at
least a third of the value for all three quark masses and that the 
results of both measurements agree within statistics. 

\begin{figure}
\begin{center}
\includegraphics[width=0.3\textwidth,angle=-90,clip]{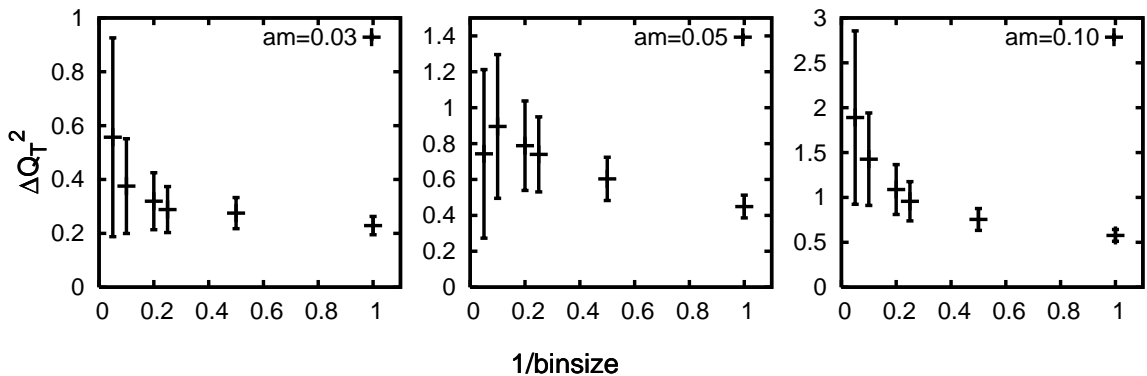}
\end{center}
\caption{
Uncertainty in measurement of $\langle Q_T^2\rangle$ as a function of bin size
(in units of two four trajectories).
(a) $am_q=0.03$, (b) $am_q=0.05$, (c) $am_q=0.10$.
\label{fig:errorvsbinsize}
}
\end{figure}

In order to get a second estimate for the error of our observables, we used jackknife binning
on data which has already been taken on every 4th trajectory only. The resulting error
as a function of block size is given in Fig.~\ref{fig:errorvsbinsize}. There is no clear plateau
of the error, again making reliable error estimates problematic.

This is as good as we can do. A better analysis needs a significantly larger statistics. With
the current ensemble, we ultimately cannot even be sure that the topological charge is thermalized.
Still, we can see that the algorithm captures the basic physics correctly and the topological charge
is significantly suppressed with smaller quark mass.
\begin{table}
\begin{tabular}{p{1cm}|p{2cm}p{2cm}p{2cm}p{2cm}}
 $m_q$  & $\langle Q \rangle $ & $\tau_{\rm int}$ &  $\langle(Q-\langle Q \rangle)^2\rangle$ & $\langle Q^2 \rangle$   \\
\hline
0.03   & -0.5(3) & 14(6) & 1.2(4)   & 1.5(5) \\
0.05   & -1.0(4) & 16(7) & 2.3(1.0) & 3.3(1.3) \\
0.10   & -1.3(6) & 27(13) & 4.0(1.5)& 5.9(2.2) \\
\end{tabular}
\caption{The average topological charge $\langle Q \rangle$, the associated 
integrated auto-correlation time $\tau_{\rm int}$,
 and $\langle(Q-\langle Q \rangle)^2\rangle$.\label{fig:AC}}
\end{table}

\begin{figure}
\begin{center}
\includegraphics[width=0.4\textwidth,angle=-90,clip]{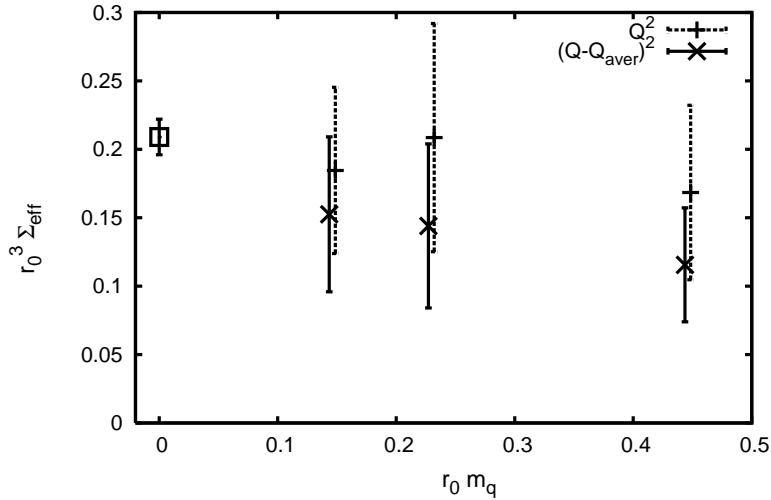}
\end{center}
\caption{
$\Sigma_{\rm eff}= Z_S \chi_T N_f/m_q$ vs. quark mass in units of $r_0$. The square at $m_q=0$ 
denotes result for  $\Sigma$  from our previous analysis\cite{DeGrand:2007tm}. 
We compare the two definitions of $\chi_T=\langle Q^2\rangle /V$ and
 $\chi_T = \langle (Q-\bar Q)^2\rangle /V$.
\label{fig:slope}
}
\end{figure}

\begin{figure}
\begin{center}
\includegraphics[width=0.8\textwidth,clip]{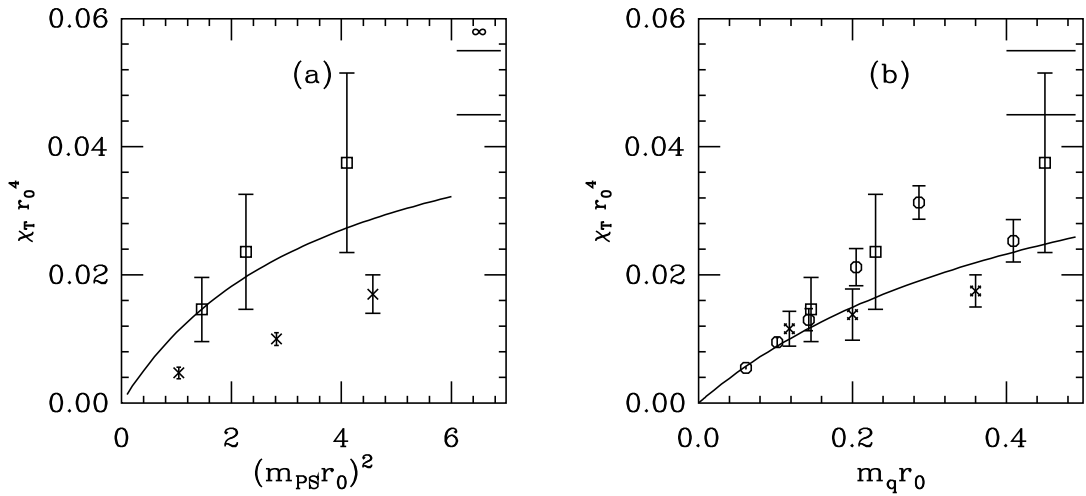}
\end{center}
\caption{Two comparisons of data: (a)
$\chi_T r_0^4$ vs $(m_{PS}r_0)^2$ with two dynamical flavors.
 Squares, our data (using $\langle Q^2\rangle$), crosses, results of
Egri, {\it et al.} \protect{\cite{Egri:2005cx}} on $8^4$ lattices.
The line is the  formula of Ref. \protect{\ref{eq:durr}}.
(b) $\chi_T r_0^4$ vs $m_q(\overline{MS})r_0$. Again, the squares are our data, while the
fancy crosses are our data from small lattices (\protect{\cite{DeGrand:2005vb}}.
Octagons are the data of  S.~Aoki, {\it et al.} \protect{\cite{Aoki:2007pw}}.
The bars show a typical determination of the quenched topological susceptibility
(from (\protect{\cite{DeGrand:2005vb}}).
\label{fig:chir04vsmpisq}
}
\end{figure}

With this caveat, we present our calculations of $\chi_T$ in 
Figs. \ref{fig:slope} and \ref{fig:chir04vsmpisq}. When the abscissa is the quark mass,
we mean the $\overline{MS}$ quark mass at a scale $\mu=2$ GeV, computed using the
RI-MOM scheme\cite{Martinelli:1994ty}
  as implemented in Ref. \cite{DeGrand:2007tm}: $Z_S=1/Z_m=0.76(3)$.
The main result is  shown in Fig.~\ref{fig:slope} where we give $r_0^3 \Sigma_{\rm eff}=
r_0^3 Z_S N_f \chi_T / m_q$ as a function of the quark mass. The results agree with the 
constant expected from Eq.~\ref{eq:lschpt}. Also the two definitions of $\chi_T$ give compatible
results. We compare with our determination of
 $r_0 \Sigma (\overline{MS},\mu=2 \ {\rm GeV})^{1/3} = 0.594(13)$.
from  Ref.~\cite{DeGrand:2007tm}, where we 
extracted $\Sigma$ from the response of the spectrum of the (valence) Dirac operator on 
an imaginary chemical potential. The two determinations are consistent.

Fig. \ref{fig:chir04vsmpisq} compares our results to other recent two-flavor
 calculations using
overlap fermions. We take the $\langle Q^2\rangle$ definition of $\chi_T$ for these plots.
We have made separate plots with abscissas of $r_0 m_q$ 
and $(r_0 m_{PS})^2$,
since previously published groups typically present their results in only one way.
All simulations presented in those plots are plagued from statistical and systematic 
uncertainties. The calculation by Egri {\it et al.} \cite{Egri:2005cx}
 was performed on very coarse lattices whereas
Aoki {\it et al.} \cite{Aoki:2007pw}
 rely on particular finite volume effects in two-point functions to 
extract $\chi_T$.
Still, given the statistical errors, there is remarkable agreement among all groups.

\begin{figure}
\begin{center}
\includegraphics[width=0.5\textwidth,clip]{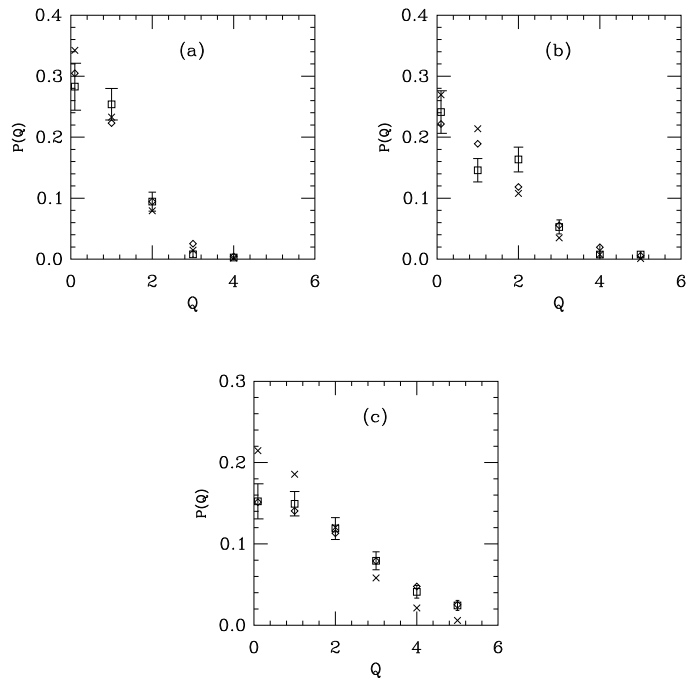}
\end{center}
\caption{
\label{fig:qvrmt}
$P(Q)$ vs $Q$ comparing  data to the ``granular''   probability (crosses), Eq.~\protect{\ref{eq:prmt}}
and the pure RMT probability (diamonds), Eq.~\protect{\ref{eq:pgran}}.
(a) $am_q=0.03$, (b) $am_q=0.05$, (c) $am_q=0.10$.
}
\end{figure}

Since the topological charge can vary during our simulations, we can compare its
probability distribution to theory and models. To do this, we symmetrize the data in
$Q$ vs $-Q$, and plot it in Fig. \ref{fig:qvrmt}.
We compare to expectations from two models.
The first is just the epsilon-regime partition function, which in a sector 
of winding number $\nu$ is
\bee
Z^{RMT}(m)= I_\nu(z)-I_{\nu+1}(z)I_{\nu-1}(z)
\label{eq:prmt}
\ee
where $z=m_q \Sigma V$ (in finite volume, $\Sigma$ is rescaled to $\Sigma_L$
with $\Sigma_L/\Sigma= 1 + (3/2)0.1405/(F^2 \sqrt{V})$ \cite{Gasser:1986vb}).
The second prediction, called the ``granular'' partition function by D\"urr\cite{Durr:2001ty},
takes
\bee
Z=Z^{RMT}(m) Z_q(\nu)
\label{eq:pgran}
\ee
where $Z_q$ is a partition function motivated by the instanton liquid:
\bee
Z_q(\nu)= \frac{1}{\sqrt{2\pi \sigma^2}}\exp(-\frac{\nu^2}{2\sigma^2})
\ee
and $\sigma^2$ is the mean-squared topological charge from a quenched simulation 
in the same volume $V$. (See also Ref. \cite{Verbaarschot:2000dy}.)
 We can infer this number from the quenched topological
susceptibility. Fig. \ref{fig:qvrmt} shows what we saw. In  it, we took
a quenched susceptibility of $\chi_Q r_0^4=0.05$ and assumed that $r_0=3.5$,
a nominal value for our simulations. Eq.~\ref{eq:prmt} seems to reproduce the data better
than Eq.~\ref{eq:pgran}.

Let us finally comment on the performance of our algorithmic setup with respect to
its ability to change topological sector.
We can quantify the tunneling rate by looking at portions of the data stream with the same
simulation parameters. For example, with three pseudofermions, the $am_q=0.03$
data set (with trajectory length 1) had a mean time between tunneling of about 8.4 trajectories.
The corresponding number at $am_q=0.05$ is 3.3. The ratio is essentially the inverse of the ratio
of squared masses ($8.4/3.3=2.5$; $(0.05/0.03)^2=2.8$). We observed this scaling
in Ref. \cite{DeGrand:2005vb}. Scaling the tunneling rate with three pseudofermions 
from $am_q=0.03$ to $am_q=0.1$,
we would expect to tunnel every 0.75 trajectories. We see a rate of about one tunnel
per 3.6 trajectories  with two pseudofermions.
The decrease of tunneling rate with smaller number of pseudofermions is also
 expected \cite{DeGrand:2006ws}.

\section{Conclusions}
The computation of the topological susceptibility in full QCD with 
chiral fermions is a challenging task. Even though the measurement
via the index theorem is easy, getting sufficient statistics for 
a reliable analysis is not. We optimized our algorithmic 
setup for good tunneling rate. However,  we still observed very long range
fluctuations in the topological charge. Those fluctuations might be
longer than our simulation time.
New ideas \cite{Cundy:2007dp} have recently been developed 
which might improve this situation.
Still, we have demonstrated
the suppression of the topological charge as the dynamical fermion mass is decreased toward
 the chiral limit. It
is consistent with the expectation from chiral perturbation theory.

\section*{Acknowledgments}

This work was supported in part by the US Department of Energy and by the Deutsche
Forschungsgemeinschaft in the SFB/TR 09.
We thank the DESY computer
center in Zeuthen for essential computer
resources and support.
We would like to thank N. Christ, Y. Shamir, and B. Svetitsky for conversations 
and correspondence
about the ``P+A trick.'' We also acknowledge instructive correspondence with
 S. D\"urr and S. Sharpe.

\end{document}